\documentstyle[11pt]{article}
\normalbaselines
\newcounter{eqnletter}[equation]
\catcode`\@=11
\@addtoreset{equation}{section}   
\catcode`\@=12

\begin{document}

\begin{center}

{\LARGE\bf Quasi-Exact Solvability in Local Field Theory}

\vskip 1cm

{\large {\bf A.G. Ushveridze} }\footnote{This work was partially
supported by DFG grant N 436 POL 113/77/0 (S)}

\vskip 0.1 cm

Department of Theoretical Physics, University of Lodz,\\
Pomorska str. 149/153, 90-236 Lodz, Poland\footnote{E-mail
address: alexush@mvii.uni.lodz.pl and alexush@krysia.uni.lodz.pl} \\
and\\
Institute of Theoretical Physics, Technical University of Clausthal,\\
Arnold Sommerfeld str. 6, 38678 Clausthal-Zellerfeld, Germany\\

\end{center}
\vspace{1 cm}
\begin{abstract}

The quantum mechanical concept of quasi-exact solvability is
based on the idea of partial algebraizability of spectral
problem. This concept is not directly extendable to the systems with
infinite number of degrees of freedom. For such systems a
new concept based on the partial Bethe Ansatz solvability
is proposed. In present paper we demonstrate the constructivity
of this concept and formulate a simple method for building
quasi-exactly solvable fild theoretical models on a
one-dimensional lattice. The method automatically leads to local models
described by hermitian hamiltonians.

\end{abstract}

\newpage

\section{Introduction}

Quasi-exactly solvable quantum models are
distinguished by the fact that they can be solved exactly
only for some limited parts of the spectrum but not for the
whole spectrum. The initial motivation for looking for such models was
associated with a natural desire of physicists and mathematicians to extend
the set of exactly solvable Schr\"odinger type equations by
relaxing the usual requirements to exact solvability. It was
quite obvious that the problem of finding models
with only a few number of exactly constructable states
should be simlper than the problem of finding new
models admitting exact solution for all the spectrum. 
This led to a natural conclusion that the set of quasi-exactly solvable 
models should be wider than that of exactly solvable ones.
The first non-trivial examples of quasi-exactly solvable models
in non-relativistic quantum mechanics
constructed in the middle of eighteens (see e.g.
reviews \cite{shif1, ush1, shif2} and book \cite{ushbook}) clearly
demonstrated that this conclusion is true. Remember, for
example, that in the class of one-dimensional models with
polynomial potentials there is only one exactly
solvable model --- the simple harmonic oscillator. 
At the same time, the number of quasi-exactly solvable
models belonging to this class is infinitely large \cite{ushbook}.

\medskip
Of course, it would be extremely important
to try to realize the idea of quasi-exact solvability in
field theory or in other systems with an infinite number of
degrees of freedom. It is however clear that the quantum
mechanical concept of quasi-exact solvability based on the
idea of a partial algebraizability of spectral problem (see
e.g. \cite{shif1}) is not directly extendable to the field theoretical 
case\footnote{Remember that a quantum problem is called
partially algebraizable if its hamiltonian admits an 
explicitly constructable finite-dimensional invariant subspace
of Hilbert space. Then a certain finite part of its 
spectrum can be constructed algebraically.}. 
The main difficulty lies in the fact that the 
physically interesting examples of quasi-exactly solvable field 
theoretical models should neccessarily contain an infinite 
number of explicitly constructable states, forming, for example, some
branches of excitations. However, the construction of such
branches cannot, generally, be qualified as an algebraic problem. 
This means that the first thing which we need is to
elaborate a constructive concept of quasi-exact solvability 
for systems with infinitely many degrees of freedom.
The following reasonings show how to do this.

\medskip
First note that the equations of mathematical physics can conventionally be
divided into three large classes distinguished from each
other by the level of their complexity. These are:

\medskip
1) the equations for a {\it finie set of numbers} (level 1)

\medskip
2) the equations for an {\it infinite set of numbers} or
{\it finite set of functions} (level 2),

\medskip
3) the equations for an {\it infinite set of functions} or
{\it finite set of functionals} (level 3). 

\medskip
In principle, this list can be continued further. Sometimes it is 
possible to reduce a problem of the $n$th level of complexity to that 
(or those) of the $(n-1)$th level.
The problems admitting such a reduction are usually called 
{\it exactly solvable}. Consider two particular cases of
this general "definition".

\medskip
1. We usually call a quantum mechanical
model {\it exactly solvable} if each its energy level and
corresponding wavefunction admits a purely {\it algebraic} construction.
This does not contradict the general "definition"
since the Schr\"odinger equation in non-relativistic 
quantum mechanics is an equation for a function (wavefunction) and thus 
the standard level of its complexity is two. At the same time,  
the level of complexity of all algebraic problems is one.

\medskip
2. We usually call a field theoretical model 
{\it exactly solvable} if the problem of construction of all its energy 
levels and correspondong wave functionals is reduced to the problem 
of solving the so-called Bethe ansatz equations (as a most
recent review see e.g. \cite{fad}).  
This is again in full accordance with the general "definition" since
the Schr\"odinger equation in field theory is an equation for a 
functional (wave functional) and therefore the level of its complexity 
is three. At the same time, the Bethe ansatz equations have 
usually the form of integral equations and therefore belong 
the second level of complexity.

\medskip
So, we see that in quantum mechanics and field theory the term "exact
solvability" has different meaning. In quantum mechanics it
means the algebraic solvability, while in quantum field theory it means 
the Bethe ansatz solvability.

\medskip
Let us now remember that the quasi-exact solvability is nothing
else than a restricted version of the exact solvability.
Restricted --- only with respect to the number of exactly computable
states. As to the level of simplification of the initial
problem, it (intuitively) is expected to be the same for both
exactly and quasi-exactly solvable systems. At least this is 
so in the case of non-relativistic quantum mechanics:
the non-relativistic quasi-exactly solvable models are defined as models 
admiting an incomplete but still {\it algebraic} solution of spectral problem.

\medskip
Taking these reasonings into account, we can propose the following extension
of the notion of quasi-exact solvability to the field theoretical case.
We shall call a field theoretical model {\it quasi-exactly solvable} if it 
admits the {\it Bethe ansatz solution} only for a certain limited 
part of the spectrum but not for the whole spectrum.

\medskip
In the present paper we demonstrate the constructivity of
this definition and present the simplest 
examples of {\it local} quasi-exactly solvable models of
field theory on a one-dimensional lattice\footnote{Note that the 
Bethe ansatz approach to the 
problem of quasi-exact solvability first appeared many years ago
in papers \cite{ushpap} where it was successfully used for building and
solving the quasi-exactly solvable problems of one- and multi-dimensional 
quantum mechanics. For further development of this approach
see refs. \cite{ush1, ushbook, wieg}.}.
Remember that the locality is a fundamental property of physically reasonable
field theories. Couriously enough, namely this property
turns out to be essential for constructing quasi-exactly
solvable field theoretical models.

\section{The basic idea}

Consider a class of quantum spin models defined on a one-dimensional 
lattice and described by the hamiltonians of the following general form:
\begin{eqnarray}
H(s_1,\ldots,s_N)= \sum_{n,m=1}^N C_{nm}\vec S_n(s_n) \vec S_m(s_m).
\label{a.1}
\end{eqnarray}
Here $\vec S_n(s_n)= \{S_{n1}(s_n),S_{n2}(s_n), S_{n3}(s_n)\}$
denote the generators of $su(2)$ algebra associated with $n$th site of 
the lattice and realizing the
$(2s_n+1)$-dimensional irreducible representation with spin $s_n$. The
corresponding representation space we denote by $W_n(s_n)$. The
matrix $C_{nm}$ is assumed to be real and symmetric,
$C_{nm}=C_{mn}$.
The hamiltonian $H(s_1,\ldots,s_N)$ is thus hermitian and acts in a
$\prod_{n=1}^N (2s_n+1)$-dimensional space
$W(s_1,\ldots,s_N)=\bigotimes_{n=1}^N W_n(s_n)$.  
The number $N$ (the length of the lattice) is assumed to be a 
free parameter which can be made arbitrarily large.

\medskip
The set of parameters $C_{nm},\ n,m=1,\ldots,N$ and $S_n,\ n=1,\ldots,N$
completely determines the model. For any values of these parameters the 
problem of solving the Schr\"odinger equation for $H(s_1,\ldots,s_N)$ in 
$W(s_1,\ldots,s_N)$ is equivalent to the problem of the diagonalization of 
a $\prod_{n=1}^N (2s_n+1)$-dimensional matrix. Generally, this can be done 
only by the help of computer (of course if $N$ is not
astronomically large). There are, however, two special cases when such
a diagonalization can be performed analytically in the
limit $N\rightarrow\infty$. Below we consider these cases separately.

\medskip
{\bf The Gaudin magnet.} Let the coefficients $C_{nm}$ be given by the formula
\begin{eqnarray}
C_{nm}=\int \frac{\rho(\lambda)}{(\lambda-a_n)(\lambda-a_m)}d\lambda
\label{a.2}
\end{eqnarray} 
in which $\rho(\lambda)$ is an arbitrary function and $a_n,\ n=1,\ldots,N$
are arbitrary real numbers. 
Such models (which are known under the generic name of
Gaudin models \cite{gau}) 
are obviously non-local, because each spin interacts with 
each other. It is known that for arbitrary collection of
spins $s_n,\ n=1,\ldots,N$ the model is completely
integrable and can be solved exactly by means of the Bethe ansatz
\cite{gau}.

\medskip
{\bf The Heisenberg magnet.} Let now the coefficients $C_{nm}$ have the form
\begin{eqnarray}
C_{nm}=2J\delta_{n,m+1}, \quad n=1,\ldots,N-1, \qquad C_{Nm}=2J\delta_{1,m}.
\label{a.3}
\end{eqnarray} 
In this case the model (which is called the Heisenberg magnet) 
becomes local, because each spin interacts only with the nearest neighbours.
It turns out that the condition of locality of the model drastically
changes its integrability properties. The Heisenberg magnet
is known to be integrable and solvable within Bethe ansatz only if
$s_n=1/2,\ n=1,\ldots,N$ \cite{fad}.

\medskip
Let us now associate with hamiltonians (\ref{a.1}) a new model,
which is formally described by the same formula,
\begin{eqnarray}
H= \sum_{n,m=1}^N C_{nm}\vec S_n \vec S_m
\label{a.4}
\end{eqnarray}
but in which the spin operators $\vec S_n$ have different meaning. Now
$\vec S_n=\{S_{n1},S_{n2},S_{n3}\}$ will denote 
spin operators realizing a certain {\it completely reducible 
representation} of algebra $su(2)$. Let 
$\sigma$ denote the set of spins characterizing the irreducible 
components of this representation. This set may be finite or infinite. 
The corresponding representation space associated
with the $n$th site of the lattice we denote by
$W_n$. For each $n$ we can write $W_n=\bigoplus_{s_n\in \sigma}
W_n(s_n)$ where $W_n(s_n)$ denote the irreducible
components of $W_n$. Thus the Hilbert space $W=\bigotimes_{n=1}^N W_n$ 
in which the operator $H$ acts can be represented as
the direct sum 
\begin{eqnarray}
W=\bigoplus_{s_1\in\sigma,\ldots,s_N\in \sigma} W(s_1,\ldots,s_N).
\label{a.5}
\end{eqnarray}
Now note that each of the spaces $W(s_1,\ldots,s_N)$ is a common
invariant subspace for the operators of the total spin 
$\vec S_n^2,\ n=1,\ldots,N$, which, obviously, commute with
each other. All these operators commute also with the
hamiltonian $H$. But this means that the spaces $W(s_1,\ldots,s_N)$
are simultaneously the invariant subspaces for $H$. 
Therefore, the spectral problem for $H$ in $W$ decomposes into an
infinite number of independent spectral problems of the type (\ref{a.1}).
After this remark let us return to cases 1 and 2 and
consider them again from the point of view of model (\ref{a.4}).
   
\medskip
{\bf The non-local case.} Let the coefficients
$C_{nm}$ in model (\ref{a.4}) be given by the formula (\ref{a.2}).
Then the hamiltonian $H$ takes a block-diagonal form
with respect to the decomposition (\ref{a.5}). The
spectral problem for each given block $H(s_1,\ldots,s_N)$,
$s_1,\ldots,s_N\in\sigma$ is the Gaudin problem. We
know that it can be solved exactly (by means of Bethe
ansatz) for any values of spins $s_1,\ldots,s_N$. But this
means that the problem of constructing the entire spectrum
of the model (\ref{a.4}) is exactly solvable.

\medskip
{\bf The local case.} Let now the coefficients $C_{nm}$
in model (\ref{a.4}) be given by the formula (\ref{a.3}).
As before, the hamiltonian $H$ takes a block
diagonal form with respect to the decomposition (\ref{a.5}).
But now {\it not any} block $H(s_1,\ldots,s_N)$,
$s_1,\ldots,s_N\in\sigma$ can be explicitly diagonalized 
by means of Bethe ansatz. If $0\not\in\sigma$, then only the Heisenberg blocks
$H(1/2,\ldots1/2)$ admit such a
diagonalization\footnote{The case when $0\in \sigma$ is a
little bit richer. We consider it separately in section 4.}. 
Now it becomes clear that the solvability properties of model 
(\ref{a.4}) are completely determined by the structure of
the set $\sigma$. Assuming that $0\not\in\sigma$, 
consider the following three cases.

\medskip 
a) The spin $s=1/2$ does not belong to the set $\sigma$. In
this case the hamiltonian $H$ does not contain the exactly
solvable blocks and the model (\ref{a.4}) is exactly non-solvable.

\medskip
b) The set $\sigma$ consists only of the spins $s=1/2$. In
this case the hamiltonian $H$ contains only the exactly
solvable Heisenberg blocks and the model (\ref{a.4}) is
exactly solvable.

\medskip
c) The set $\sigma$ contains the spin $s=1/2$ and at least
one differing spin $s\neq 1/2$. In this case only a part of
hamiltonian blocks admit explicit diagonalization so that
we deal with a typical case of quasi-exactly solvable model!

\medskip
Summarizing, one can claim that the model with hamiltonian
\begin{eqnarray}
H= 2J\sum_{n=1}^N \vec S_n \vec S_{n+1}
\label{a.6}
\end{eqnarray}
is quasi-exactly solvable, provided that the representation
in which the spin operators $\vec S_n$ act is completely
reducible and contains at least one representation of spin
$s=1/2$ and at least one representation of other spin
$s\neq 1/2$. Below we shall refer this condition to as condition of 
$1/2$-reducibility.

\medskip
Consider the simplest example. Let for any $n$ $W_n$ be a direct product
of two representation spaces of irreducible representations
with spins $s=2$ and $s=3/2$:
$W_n=W_n(2)\otimes W_n(3/2)$. This
means that the hamiltonian (\ref{a.6}) can be represented
in the form
\begin{eqnarray}
H= 2J\sum_{n=1}^N (\vec S_n(2)+\vec S_n(3/2))
(\vec S_{n+1}(2)+\vec S_{n+1}(3/2))
\label{a.61}
\end{eqnarray}
where $S_n(2)$ and $S_n(3/2)$ denote the corresponding
spin operators. For any $n$ the representation $W_n$ is completely reducible
and its decomposition in irreducible components reads: 
$W_n=W_n(1/2)\oplus W_n(3/2)\oplus W_n(5/2)\oplus W_n(7/2)$. We
see that it satisfies the condition of $1/2$-reducibility
and therefore the model (\ref{a.61}) is quasi-exactly solvable.

\medskip
Before completing this section let us 
reduce the model (\ref{a.6}) to a little more
convenient form. Taking into account the fact that the
hamiltonian ${\cal H}$ commutes with Casimir invariants $\vec
S_{n}^2$ of algebra $su(2)$ (whose spectrum in each block is trivial), 
we can conclude that the modified hamiltonians ${\cal H}=H+
\sum_{n=1}^M b_n\vec S_n^2$ with arbitrary coefficients
$b_n$, will be quasi-exactly solvable, as well. In particular,
taking $b_n=2|J|, \ n=1,\ldots, N$ we obtain a new hamiltonian
\begin{eqnarray}
{\cal H}= |J|\sum_{n=1}^N (\vec S_n \pm \vec S_{n+1})^2
\label{a.7}
\end{eqnarray}
which is bounded from below and in which $\pm\equiv\mbox{Sign}\ J$.
We see that sign `$-$' corresponds to a ferromagnet
case $J>0$ and the sign `$+$' describes the anti-ferromagnet case
$J<0$.

\section{The QES families}

It is known that the quasi-exactly solvable models of
non-relativistic quantum mechanics usually appear in the
form of infinite sequences of models which look more or
less similarly but differ from each other by the number of exactly
computable states. This number, which is usually called the
order of a quasi-exactly solvable model, is gouverned by
the set of discrete parameters in the
potential. For example, the potentials of
simplest sequence of quasi-exactly solvable sextic anharmonic oscillator 
models are parametrized by a semi-integer parameter $s$ and
read $V(x)=x^6-(8s+3)x^2$. For any given $s=0,1/2,1,3/2,\ldots$ the
model admits $2s+1$ algebraically constructable solutions.
It is naturally to call this phenomenon the phenomenon of quantization of
potential.

\medskip
Now it is naturally to ask, if there is any analogue of the
quantization of potential in the field theoretical case?
Or, more concretely, if the model we constructed in the
previous section is a member of a certain infinite sequence of
similar models?

\medskip
The answer to this question is positive.
In order to show this it is sufficient to remember the well known result 
in the theory of completely integrable quantum systems
about the generalization of completely integrable Heisenberg 
chain to the case of arbitrary spin. This result is
highly non-trivial because the naive substitution of $1/2$ spin 
operators $S_n(1/2)$ by the arbitrary spin operators $S_n(s_n)$
does lead to integrable model. The correct generalization
includes the change of potential describing the interaction
of neighbouring spins. The form of the generalized 
hamiltonian is
\begin{eqnarray}
H_P(s,\ldots,s)= J\sum_{n,m=1}^N P_{2s}[\vec S_n(s) \vec S_{n+1}(s)]
\label{a.10}
\end{eqnarray}
where $\vec S_n(s)$ are the operators of spin $s$ and
$P_{2s}[t]$ is a very specific polynomial of degree $2s$
defined by the formula
\begin{eqnarray}
P_{2s}[t]=2\sum_{i=0}^{2s}\left(\sum_{j=i+1}^{2s}\frac{1}{j}\right)
\prod_{k=0, k\neq i}^{2s}\frac{t-t_k}{t_i-t_k}
\label{a.11}
\end{eqnarray}
with
\begin{eqnarray}
t_k=\frac{1}{2}k(k+1)-l(l+1).
\label{a.12}
\end{eqnarray}
It turns out that if the form of polynomial $P_{2s}[t]$
differs from that given by formula (\ref{a.11}) then the spin system
(\ref{a.10}) becomes non-integrable and exactly non-solvable.

\medskip
Repeating the reasonings of section 2, let us now consider
the hamiltonian
\begin{eqnarray}
H_P = J\sum_{n=1}^N P_{2s}[\vec S_n \vec S_{n+1}]
\label{a.13}
\end{eqnarray}
in which, as before, $\vec S_n$ denote the spin operators
acting in a certain completely reducible representation of
algebra $su(2)$. In full analogy with the previous case the
model (\ref{a.13}) is quasi-exactly solvable if 
this representation contains at least one irreducible representation 
of spin $s$ and at least one representation of spin $s'\neq s$.
Such representations we shall call $s$-reducible.

\medskip
Assume now that the completely reducible representation in which the operators
$\vec S_n$ act contains {\it all}
irreducible representations of spins $s_n=0,1/2,1,3/2,\ldots$.
There are many ways for building such a representation. One
of the simples ways is to write
\begin{eqnarray}
S_{n1}&=&\frac{1}{2}(a_n^+b_n+b_n^+a_n),\nonumber\\
S_{n2}&=&\frac{1}{2i}(a_n^+b_n-b_n^+a_n),\label{a.8}\\
S_{n3}&=&\frac{1}{2}(a_n^+a_n-b_n^+b_n),\nonumber
\end{eqnarray}
where $a_n, a_n^+$, and $b_n, b_n^+$ denote two
independent groups of hermite conjugated annihilation and
creation bosonic operators obeying the Heisenberg commutation relations
$[a_n,a_n^+]=1$ and $[b_n,b_n^+]=1$ (the commutators
between the $a$- and $b$- operators and also
between the operators associated with different sites are zero). 
Indeed, it is obvious that the operators defined by formulas (\ref{a.8}) 
are hermitian and obey the standard commutation relations of $su(2)$ algebra.
Moreover, if we denote by $|M_{a_n},M_{b_n}\rangle$ the
states with given numbers of $a$-bosonic and
$b$-bosonic quants, then the set of all 
such states with $M_{a_n}+M_{b_n}=2s_n$ will form the
basis of the $(2s_n+1)$-dimensional 
irreducible representation of algebra $su(2)$
with spin $S_n$. This means that the representation defined
by formulas (\ref{a.8}) is completely reducible and 
contains all finite-dimensional irreducible representations 
of algebra $su(2)$ with multiplicity 1. This means that
for any given $s$ this representation is $s$-reducible,
which, in turn, implies the quasi-exact solvability of
model (\ref{a.13}) for any given $s$. 

\medskip
The quasi-exactly solvable models constructed
above describe the interaction of two bosonic fields on a lattice.
So we see that these models form an infinite sequence.
The role of a semi-integer parameter $s$ quantizing the potentials
of quasi-exactly solvable sextic anharmonic oscillator is
now played by the function $P_{2s}[t]$. This is quite
natural because for field theoretical problems (having
the third level of complexity) the functions play the same
role as numbers for quantum mechanical problems (whose
level of complexity is two).

\medskip
It is not difficult to construct for any $s$ the analogs of
modified hamiltonians (\ref{a.7}). To do this, let us introduce
the new polynomials $Q_{2s}[t]$ related to the old ones, $P_{2s}[t]$,
by the formula:
\begin{eqnarray}
Q_{2s}[2s(s+1)\pm 2t]=P_{2s}[t]
\label{a.14}
\end{eqnarray}
Repeating the reasonongs of section 2 we can easily
conclude that the models with hamiltonians
\begin{eqnarray}
{\cal H}_Q = \sum_{n}^N Q_{2s}[(\vec S_n \pm \vec S_{n+1})^2]
\label{a.15}
\end{eqnarray} 
will be quasi-exactly solvable if the model (\ref{a.13}) is.

\medskip
It is worth stressing that for integer spins the polynomial
$Q_{2s}[t]$ is even and, according to formulas (\ref{a.11}),
(\ref{a.12}) and (\ref{a.14}) has positive coefficient at the
leading term. Therefore, for $J>0$ the spectrum of such models is
bounded from below irrespective of the sign `$\pm$'. If the
spin is semi-integer, then the leading term of polynomial
$Q_{2s}[t]$ is positive for sign `$+$' and negative for
sign `$-$'. Therefore, the spectrum of the model will be
bounded from below if $\mbox{Sign}\ J=\pm$.

\section{Possible generalizations}

The method discussed above admits many natural generalizations.
We divide them into three groups.
The first group concerns the form of the initial
hamiltonian. One can consider the following important subcases:

\medskip
{\it 1. The anisotropic case}. In this paper we considered only 
the isotropic $XXX$-magnets invariant under global $su(2)$-rotations.
However, all the reasonings given above remain valid for anisotropic
$XXY$ magnets and their generalizations for higher spins
constructed in ref. \cite{tar}.

\medskip
{\it 2. The case of higher integrals of motion}. The $XXX$ and $XXY$
magnets admit an infinite set of integrals of motion.
These integrals can be considered as hamiltonians 
of local spin chains with different number of interacting spins.
All the reasonongs of the present paper can be repeated for
these hamiltonians.

\medskip
{\it 3. The case of higher Lie algebras}. 
Instead of local $su(2)$ magnets one can consider their
generalizations for arbitrary simple Lie algebras. 

\medskip
The second group of generalizations concerns the
realization of completely reducible unitary representations
of Lie algebras. The two examples of such a realization
considered in sections 2 and 3 do not exhaust all the existing
possibilities whose variety is very large. Below we
consider some of them restricting ourselves to the
simplest $su(2)$ case.

\medskip
{\it 1. The spin realization}. In this realization the
extended spin operators have the form
\begin{eqnarray}
\vec S_n=\vec S_n(s^{(1)}) + \ldots + \vec S_n(s^{(K)})
\label{a.16}
\end{eqnarray} 
where $s^{(1)},\ldots,s^{(K)}$ are arbitrary spins. 
This  representation is completely reducible and contains 
all representations with spins lying (with spacing 1)
between  $s_{\min}=\min |s^{(1)}\pm \ldots\pm s^{(K)}|$ and
$s_{\max}=\max |s^{(1)}\pm \ldots\pm s^{(K)}|$. The
resulting quasi-exactly solvable models form a finite
family with $s_{\max}-s_{\min}+1$ members. In formula (\ref{a.16})
the extension is assumed to be homogeneous, i.e. the
spins $s^{(1)},\ldots, s^{(K)}$ are assumed to be independent on $n$. 
However, equally well we could consider the
inhomogeneous case.

\medskip
{\it 2. The bosonic realizations}. There are several ways of
expressing the generators of $su(2)$ algebra via the
generators of Heisenberg algebra. One of such ways we
considered in section 3 (see formula (\ref{a.8})).
Another way can be based on the following realization of
spin operators
\begin{eqnarray}
S_{n1}&=&p_{n2}q_{n3}-p_{n3}q_{n2},\nonumber\\
S_{n2}&=&p_{n3}q_{n1}-p_{n1}q_{n3},\label{a.17}\\
S_{n3}&=&p_{n1}q_{n2}-p_{n2}q_{n1}.\nonumber
\end{eqnarray} 
Here $\vec q_n$  are the components of a
certain (real) bosonic vector field and
$\vec p_n$ denote the components of the corresponding
(real) generalized momentum. These components satisfy the
Heisenberg commutation relations $[p_{ni},q_{mk}]=i\delta_{nm}\delta_{ik}$.
The representation in which the operators (\ref{a.17}) act
is infinite-dimensional and completely reducible,
but, in contrast with the representation defined by formula
(\ref{a.8}),
it does not contain any irreducible components with half integer 
values of $s$. As before, the realization (\ref{a.17}) leads to 
an infinite sequence of models of the type (\ref{a.15}),
but now these models  are quasi-exactly solvable only for 
integer values of $s$. As to the models with half integer values of $s$,
they all become exactly non-solvable.

\medskip
{\it 3. The fermionic realizations}. Along with the bosonic
realizations considered above, we can consider
the fermionic ones. The simplest fermionic realization
looks like the bosonic one given by formula (\ref{a.8})
and reads
\begin{eqnarray}
S_{n1}&=&\frac{1}{2}(f_n^+g_n+g_n^+f_n),\nonumber\\
S_{n2}&=&\frac{1}{2i}(f_n^+g_n-g_n^+f_n),\label{a.19}\\
S_{n3}&=&\frac{1}{2}(f_n^+f_n-g_n^+g_n).\nonumber
\end{eqnarray}
Here the operators $f_n^\pm$ and $g_n^\pm$ satisfy
the Heisenberg {\it anti-commutation} relations 
$\{f_n,f_n^+\}=1$, $\{g_n,g_n^+\}=1$
(the anti-commutators between the $f$- and $g$-operators and also
between the operators associated with different sites are zero). 
As before, the operators defined by formulas (\ref{a.19}) are hermitian 
and obey the standard commutation relations of $su(2)$ algebra.
The representation in which the operators (\ref{a.19}) act 
is four-dimensional and decomposes into three
irreducible representation with spins $1/2$, $0$ and $0$. 
This means that it does not satisfy the condition of $s$-reducibility
and thus cannot lead to any quasi-exactly solvable model.
In order to improve the situation one can consider a tensor
product of several representations of the type (\ref{a.19}). Then the
resulting (composite) representation will satisfy the
$s$-reducibility condition for a certain finite set of
spins. This will lead to a finite family of quasi-exactly
solvable models describing the interaction of several
fermionic fields on a lattice. 

\medskip
{\it 4. The mixed cases}. The mixed cases appear when
the type of the realization (spin, bosonic, fermionic)
changes from site to site.  The mixed realizations may lead to
quasi-exactly solvable models describing the interaction of
different fields, say, fermionic and bosonic fields.

\medskip
The third group of generalizations concerns the 
additional solutions which, up to now, we did not discuss in this paper.
In order to explain the appearence of these solutions in 
the simplest $su(2)$ case, let us consider again the
spectral problem for hamiltonian (\ref{a.13})
discussed in section 3. Remember that this problem
decomposes into a set of independent spectral problems
for the spin hamiltonians
\begin{eqnarray}
H_P(s_1,\ldots,s_N)= J\sum_{n,m=1}^N P_{2s}[\vec S_n(s_n) 
\vec S_{n+1}(s_{n+1})].
\label{a.20}
\end{eqnarray}
Remember also that the hamiltonian with $s_n=s$ is
explicitly diagonalizable within Bethe ansatz which gives
us a part of explicit solutions of the initial problem (\ref{a.13}). 
The additional solutions appear when the
spin operators $\vec S_n$ in formula (\ref{a.13}) realize a
completely reducible representation of algebra $su(2)$ containing a
one-dimensional irreducible component with spin zero. 
Note that the generators of this one-dimensional component vanish: 
$\vec S_n(0)=\vec 0$. 
For this reason, those of models (\ref{a.20}) which contain
zero spins become equivalent to a system of several disconnected 
inhomogeneous spin chains with open
ends.  It is reasonable to distinguish between the
following opposite cases.

\medskip
1. The open sub-chains are homogeneous ($s_n=s$) and very long.
In this case the boundary effects become negligibly small
and all open sub-chains can be approximated by periodic ones.
The latter are however exactly solvable and this extends
the set of explicit solutions of model (\ref{a.13}).

\medskip
2. The open sub-chains are inhomogeneous ($s_n\neq s$) and short.
Then the diagonalization can be performed algebraically. Note that 
algebraic diagonalizability of hamiltonians of short chains
does not require the condition of Bethe ansatz solvability. This means 
that the condition of homogeneity of short chains becomes unneccessary.
They equally well may consist of spins differing of $s$.

\medskip
So we see that the presence of zero spin components in a
completely reducible representation of a Lie algebra
considerably extends the set of states admitting explicit solutions.

\section{Concluding remarks}

From the above consideration it follows that there are
remarkable parallels between the methods of constructing quasi-exactly
solvable problems in quantum mechanics (QM) and in field
theory (FT) on a lattice. Both methods start with a certain 
spin system. This is a quantum top in QM case 
and infinite spin chain in FT case. Both systems are
exactly solvable. This is the algebraic solvability in QM
case and Bethe ansatz solvability in FT case. In both
systems the spins realize a certain finite-dimensional and
irreducible matrix representation $T_{fin}$  of algebra
$su(2)$. In both cases one replaces the spin hamiltonian by
a new extended hamiltonian which formally has the same form
as the initial one but in which the spin operators have
different meaning. Now they realize a certain infinite-dimensional 
and reducible representation $T_{inf}$ of algebra $su(2)$.
In both cases however this representation contains  $T_{fin}$ as 
an irreducible component. This finally leads to the
quasi-exact solvability of the extended hamiltonian.
As we noted above, both the QM and FT quasi-exactly solvable
models appear in the form of infinite sequences.

It is worth stressing, however, that along with many
common features there is a drastical difference between the 
quantum mechanical and field theoretical quasi-exactly
solvable models. This difference concerns the principle of
building the infinite-dimensional representation spaces $T_{inf}$.
Indeed, in QM case there is no neccessity of changing the form of the
initial spin hamiltonian if one changes the representation
in which the spin acts. For any finite-dimensional representation
$T_{fin}$ the quantum top is exactly (algebraically) solvable. This
means that the only way of reducing this hamiltonian to a 
quasi-exactly solvable form is to use such an infinite
dimensional and reducible representation $T_{inf}$ which contains only 
a finite number of finite-dimensional irreducible components $T_{fin}$.
Such representations do actually exist but they are
non-unitary and the corresponding extended spin operators
become non-hermitian. This produces considerable
difficulties in constructing hermitian quasi-exactly
solvable models in quantum mechanics.

In FT case the situation is different. Now the Bethe ansatz
solvability of the initial spin hamiltonian strongly
depends on the representation in which the spins act. For
this reason the representation $T_{inf}$ may now contain an infinite
number of finite-dimensional irreducible components. At any
rate the extended hamiltonian will have only one invariant
subspace in which it will be exactly solvable. 
This means that we obtain a great freedom in chosing the
representation $T_{inf}$. There are many unitary
representations of such a sort which automatically lead to
hermitian quasi-exactly solvable models of field theory.
In this sense the proposed procedure of building
quasi-exactly solvable problems in field 
theory is conceptually simpler than that in non-relativistic 
quantum mechanics.

\section{Acknowledgements}

I take the opportunity to thank Pavel Wiegmann for
interesting discussions in Toronto.

\end{document}